# Reflection and refraction properties of laser-driven 2D quantum well: Analogy with Photonic Time Crystal

Igor V. Smetanin,[1*] and Alexander V. Uskov[1]

[1]P. N. Lebedev Physical Institute of Russian Academy of Sciences, Leninsky pr. 53, Moscow, 119991, Russia

**ABSTRACT**. It has been demonstrated that a quantum well with optically excited, homogeneous oscillations of a two-dimensional (2D) electron plasma behaves as a photonic time crystal when it scatters an obliquely incident, weak probe electromagnetic wave. The hydrodynamic approximation is used to describe self-consistently the interaction between the probe wave and the 2D plasma. Such a quantum well becomes a polychromatic source and to the first order in perturbation reflected and transmitted radiation reveals as non-shifted in frequency principal components along with up- and down-shifted in frequency satellites. The downshifted satellites are backward propagating when the frequency of the probe wave is less than that of the plasma oscillations. We found the condition when the downshifted components are the surface waves. For the s-polarized probe wave, we found that the laser-driven quantum well can support propagation of TE surface wave, which would not normally be permitted to propagate freely. The amplitudes of the reflected and transmitted components are determined for the principal components and the up- and down-shifted satellites.

## I. INTRODUCTION.

The concept of a photonic time crystal (PTC), which is a medium that is spatially uniform and has a refractive index that oscillates periodically over time, is now widely discussed as a prospective object in the field of photonics with some rather unusual optical properties [1-3]. PTCs are considered to be the counterpart of conventional photonic crystals (PC), which are optical (nano-) structure in which the refractive index exhibits spatial periodicity. The spatial periodicity of the refractive index in PC affects the propagation of light in the same way that periodic potential in natural crystals affects the motion of electrons. Namely, the spatial periodicity of the refractive index gives rise to the formation of photon energy band gaps in PC, i.e. intervals of frequencies at which the propagation of light in PC is prohibited. By contrast, in PTC, changes in the refractive index over time result in photon momentum band gaps, i.e. intervals of wavenumbers at which the propagation of light in PTC is prohibited [1-3]. In fact, the band structure of PTCs is similar to that of spatial photonic crystals but rotated within the ω–k plane [4, 5]. Unlike conventional PCs, this bandgap is characterized by exponential energy amplification rather than wave suppression [6].

PTCs provide us with a variety of novel light generation and amplification phenomena. These include a new mechanism of gain in time-dependent media [7], vacuum amplification effects in anisotropic temporal boundaries including inhibition of photon production along specific directions [8], amplified light emission from quantum emitters [6] and free electrons through subluminal Cherenkov radiation [9,10], superluminal momentum-gap solitons [11] and incandescent light sources that are not constrained within the black-body spectrum [12].

The peculiarities of reflection and refraction in PTC offer novel unprecedented opportunities for light manipulation , including the creation of compact, low-energy, non-reciprocal devices [13], novel anti-reflection coatings [14, 15], the realization of various logic procedures using PTCs with energy loss [16], quantum state frequency shifting and ultra-fast switching without thermal noise amplification [15], the temporal equivalent of the Brewster angle [17], the inverse (anti-Newtonian) prisms [18] and temporal aiming [19].

The propagation of light through slabs with a dielectric function, changing in time periodically, has been carefully studied for normal incidence [20–23]. It has been demonstrated [20] that in the case of an incident monochromatic wave the slab behaves as a polychromatic source, generating multiple harmonics ω −nΩ, ω is the frequency of the incident wave, Ω is the modulation frequency of the dielectric function, n is an integer. Manley-Raw relationship does not satisfy for these harmonics because energy conservation is violated in time modulated media (energy is pumped into the system through the modulation). The dynamic-periodic slab response can exhibit parametric resonances at the incident wave frequency, provided that the relation ω = mΩ/2 is satisfied where m = 1,3,5... [21, 22]. These results were extended to the case of a quasi-monochromatic pulse in [23]. The scattering of obliquely incident electromagnetic waves from periodically space-time modulated slabs is investigated in [24].

The experimental realization of PTCs is challenging, particularly in the optical domain since the modulation must be at optical speeds [25]. Recently, the experimental time-reflection of microwaves by a fast, optically controlled time-boundary at the highest frequency ever observed (0.59 GHz) has been reported recently [26]. The prospect of PTCs operating in the optical domain is now associated, in particular, with epsilon-near-zero metamaterials [27].

In this letter, we examine the reflection and refraction properties of a laser-driven two-dimensional quantum well (QW), which exhibits a clear analogy with the PTC in the THz-to-optical frequency domain. We explore the approach which has been previously used to develop the method of coherent resonance excitation of THz plasma oscillations in a 2D QW by two-color laser radiation due to photo-induced generation of electron-hole pairs [28]. Below, we consider the case in which two driver laser beams excite homogeneous oscillations in the photo-induced electron-hole 2D plasma density at the beat frequency $\Omega$, and investigate the reflection and transmission characteristics of the obliquely incident monochromatic optical probe wave for both p- and s-polarization. The hydrodynamic approximation [29] is used to describe the self-consistent light-2D plasma interaction. First-order perturbation theory shows that the reflected and transmitted waves consist of non-shifted in frequency principal modes propagating in the forward direction at angles equal to the angle of incidence, as well as satellite modes that are shifted up and down in frequency, $\omega \pm \Omega$ ($\omega$ is the probe wave frequency). When the probe wave frequency exceeds the beat frequency, $\omega > \Omega$, the reflected and refracted satellites propagate in the forward direction at the angles close to the normal axis for an up-shifted satellite or to the surface for a down-shifted satellite, see Fig.1. In the opposite case, when the beat frequency exceeds the probe wave frequency $\Omega > \omega$, the down-shifted signal propagates backwards, as shown in Fig.1. If the difference $|\omega - \Omega|$ is enough small, the down-shifted refracted and reflected signals appear as surface waves travelling forwards or backwards depending on the sign of $\omega - \Omega$. We demonstrate that this laser-driven 2D QW supports a transverse electric (TE) surface wave in the case of incident s-polarized probe wave: Note, free TE surface waves are conventionally prohibited at a metal-dielectric interface, only the transverse-magnetic free surface waves exist. Below, we calculate the refraction and reflection coefficients for the principal components, as well as the conversion amplitudes for satellites that are shifted up or down in frequency. All these features allow us to consider the laser-driven QW as a kind of PTC suitable for experimental realization in THz-to-optical frequency domain.

## II. INTERACTION OF ELECTROMAGNETIC WAVE WITH LASER-DRIVEN 2D QW

The schematic geometry of the laser-driven two-dimensional (2D) quantum well is shown in the Fig. 1. We assume that the QW surface is located in the $yz$ plane and that the $x$-axis is normal to it. Two driving laser beams with the intensities $P_{1,2}$ and the frequencies $\omega_{1,2}$ are incident on the QW almost normally, at the sufficiently small angles $\vartheta_{1,2}$, so that $(k_1 \sin \vartheta_1 + k_2 \sin \vartheta_2) L \ll 1$, $k_{1,2} = \sqrt{\varepsilon} \omega_{1,2} / c$ are the wavenumbers, $\varepsilon$ is the dielectric permittivity of the medium in which QW is embedded, and $L$ is the characteristic QW length along the $z$-axis. Then, the interference intensity distribution in the QW plane is homogeneous and oscillates in time, $P|_{x=0} = P_1 + P_2 + 2\sqrt{P_1 P_2} \cos(\Omega t + \Delta\phi)$, at the beat frequency $\Omega = \omega_2 - \omega_1$, $\Delta\phi$ is the phase difference. Assuming that both laser frequencies exceed the QW bandgap energy, a homogeneous distribution of electron-hole pairs over the QW is achieved, with a density that oscillates in time at the beat frequency $\Omega$ [28].

We are interested in the steady-state reflection and transmission of an oblique incident electromagnetic probe wave, being s- or p- polarized. Let us first consider the case of p-polarized incident wave

$$\begin{aligned} \vec{E}_p &= E_{p0}(\hat{e}_x \sin \theta + \hat{e}_z \cos \theta) \times \\ &\times \exp[i(k \sin \theta)z - i(k \cos \theta)x - i\omega t] + c.c. \end{aligned} \quad (1)$$

Here, $E_{p0}$, $\omega$ and $k$ are the amplitude, the frequency and the wavenumber of the probe wave, respectively, $k^2 = \varepsilon \omega^2 / c^2$, $\theta$ is the angle of incidence, $\hat{e}_x$ and $\hat{e}_z$ are the unit vectors along the axes $x$ and $z$, correspondingly. It is reasonable to expect that the reflected and transmitted fields consist of two principal components that are not shifted in frequency, as well as a set of harmonics that are shifted in frequency, with frequencies given by $\omega \pm q\Omega$ and $q = 1, 2, \dots$. The principal components should have the same wavenumber along the $z$-axis: $k \sin \theta$. This inevitably means that the reflection and transmission angles for non-shifted components are equal to the angle of incidence due to the dispersion relationship.

*Contact author: smetaniniv@lebedev.ru

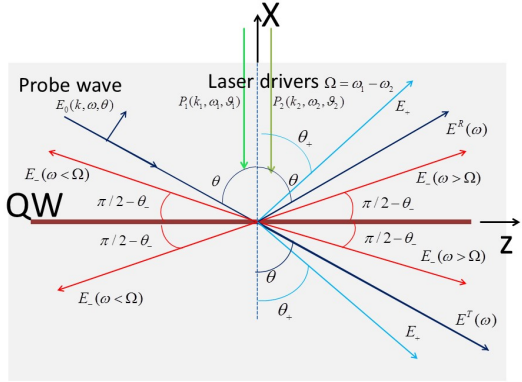

FIG 1. Diagram showing the reflected and transmitted waves scattered by a 2D quantum well (QW) driven by laser radiation. Two driver laser beams (green arrows) are incident almost normally on the 2D QW at the $x=0$ plane. The incident probe electromagnetic wave $E_0(k, \omega, \theta)$ and the principal reflected $E^R(\omega)$ and transmitted $E^T(\omega)$ components propagating at the angle of incidence $\theta$ are shown in dark blue, the up-shifted in frequency $\omega + \Omega$ satellites $E_+$ are shown as blue rays at the angles $\theta_+$, the down-shifted in frequency $|\omega - \Omega|$ satellites are shown by red rays, both forwards $E_-(\omega > \Omega)$ and backwards $E_-(\omega < \Omega)$.

Assuming the holes in QW are heavy, the induced polarization is determined by the oscillations of the 2D electron density $n(z,t)$ and the longitudinal velocity $v(z,t)$ of 2D electrons which in the steady-state regime are described by the following hydrodynamic equations [29]

$$\frac{\partial n}{\partial t} + \frac{\partial (nv)}{\partial z} = -\gamma_R(n - n_0) + \\ + 2\sigma\sqrt{P_1 P_2}\cos(\Omega t + \Delta\varphi) \qquad (2)$$

$$\frac{\partial v}{\partial t} + v\frac{\partial v}{\partial z} = -\frac{s^2}{n_0}\frac{\partial n}{\partial z} - \frac{e}{m}\tilde{E}_{slf}(z,t) - \gamma v - \\ - A\exp[ikz\sin\theta - i\omega t] + c.c.$$

Here, the external force, driving the electrons along the $z$-axis, is given by $z$-component of the total electric field of the incoming probe wave, the reflected and transmitted waves. Below we show that up- and down-shifted in frequency components are sufficiently weak, so that one can write $A = (e/m)(E_{p0} - E^R_{p\omega} + E^T_{p\omega})\cos\theta$, where $E^R_{p\omega}$ and $E^T_{p\omega}$ are the amplitudes of the reflected and transmitted principal (non-shifted in frequency) components.

*Contact author: smetaniniv@lebedev.ru

components. The steady-state 2D plasma density in QW is $n_0 = n_i + \sigma(P_1 + P_2)/\gamma_R$, where $n_i$ is the 2D plasma density in QW without laser irradiation, $\sigma$ is the rate at which electron-hole pairs are produced by light, $s$ is the speed of sound of the 2D electron gas, and $\tilde{E}_{slf}(z,t)$ is the self-consistent longitudinal electric field, generated in the 2D electron plasma due to a spatial inhomogeneity of the electron density which is induced by an external field.

We seek a solution to Eqs. (2) in the form

$$n = n_0 + n_1(t) + (\rho(t)\exp[ihz] + c.c.) \\ v(z,t) = u(t)\exp[ihz] + c.c. \qquad (3)$$

where we introduce $h = k\sin\theta$ for convenience. By substituting (3) into (2), we can show that the homogeneous steady-state oscillations of the 2D plasma density can be expressed as follows:

$$n_1(t) = n_{10}\cos(\Omega t), \quad n_{10} = 2\sigma\sqrt{P_1 P_2}(\Omega^2 + \gamma_R^2)^{-1/2}, \quad (4)$$

without loss of generality of our considerations.

Assuming the probe electromagnetic wave is weak, the linearization procedure leads to

$$\frac{\partial \rho}{\partial t} + \gamma_R\rho = -ih(n_0 + n_1)u \\ \frac{\partial u}{\partial t} + \gamma u = -i\frac{hs^2}{n_0}\rho - \frac{e}{m}E_{slf} - A\exp[-i\omega t] \qquad (5)$$

Here, the self-consistent electric field is $\tilde{E}_{slf}(z,t) = E_{slf}(t)\exp[ihz] + c.c.$, and according to the Gauss theorem [29], the amplitude of its spatial harmonic is related to the amplitude of the spatial harmonic of the 2D density as $E_{slf}(t) = i(2\pi e/\varepsilon)\rho(t)$ [29].

From Eq. (5), it is easy to see that the velocity perturbation satisfies the forced, damped Mathieu equation

$$\frac{d^2u}{dt^2} + \delta\frac{du}{dt} + \omega_0^2(1 + \alpha\cos(\Omega t))u = \\ = i(\omega + i\gamma_R)A\exp(-i\omega t) \qquad (6)$$

where $\delta = \gamma + \gamma_R$ is the damping rate, the eigen frequency and the modulation parameters are

$$\omega_0 = \left(\gamma\gamma_R + \Omega_p^2\right)^{1/2}, \quad \alpha = \frac{\Omega_p^2}{\gamma\gamma_R + \Omega_p^2}\frac{n_{10}}{n_0} \qquad (7)$$

respectively. Here, the conventional 2D plasma frequency is $\Omega_p^2 = 2\pi n_0 e^2 h / m\varepsilon$ [29]. It should be noted that Eq. (6) shows the similarity between the standard PTC and the scheme under consideration. Really, according to the Ince-Strutt diagram [30], the free oscillations of an unforced and undamped Mathieu oscillator ($A = 0$, $\delta = 0$) become unstable at frequencies close to the resonances $\omega_0 = N\Omega / 2$, $N = 1, 2, \ldots$; in other words, at the wavenumbers determined by the roots of the equation

$$h^2 s^2 + \frac{2\pi n_0 e^2}{m\varepsilon}h + \gamma\gamma_R - \frac{N^2\Omega^2}{4} = 0 \qquad (8)$$

That is, "band gaps" emerge at longitudinal wavenumbers $h$ at which the propagation of the signal wave becomes unstable, in analogy with conventional PTCs. However for the damped Mathieu oscillator ($\delta > 0$), the above instability does not emerge at sufficiently small modulation parameters, $\alpha < \delta\Omega / \omega_0^2$ at $N=1$, and the solution to the Mathieu equation remains regular in time.

We are interested in the steady-state solutions of Eqs. (6). Assuming $P_1 \approx P_2$ and small relaxation rates compared to the beat frequency, $\gamma / \Omega \ll 1$, $\gamma_R / \Omega \ll 1$, we find $\alpha \approx \gamma_R / 2\Omega \ll 1$. Thus, since $\alpha\omega_0^2 / \delta\Omega \approx (\gamma_R / \gamma)\omega_0^2 / \Omega^2 \ll 1$, one should expect that Eq. (6) has regular solution across the entire frequency range. According to Eq. (6), multiple harmonics $\omega \pm q\Omega$, $q = 1, 2, 3 \ldots$ are generated; that is to say, QW becomes a polychromatic source, analogous to the time-modulated slab considered in [20]. Below, we restrict ourselves to the first order in $\alpha$, so that the solution of the forced Mathieu equation (6) is given by

$$u(t) = iA\frac{\omega + i\gamma_R}{D_0}\Big(\exp[-\mathrm{i}\,\omega t] -$$
$$-\frac{\alpha}{2}\frac{\omega_0^2}{D_-}\exp[-\mathrm{i}(\omega - \Omega)t] - \frac{\alpha}{2}\frac{\omega_0^2}{D_+}\exp[-\mathrm{i}(\omega + \Omega)t]\Big) \qquad (9)$$

where the determinants are $D_0 = \omega_0^2 - \omega^2 - i\delta\omega$, and $D_\pm = \omega_0^2 - (\omega \pm \Omega)^2 - i\delta(\omega \pm \Omega)$.

## III. RESULTS FOR THE P-POLARIZED WAVE

*Contact author: smetaninv@lebedev.ru

Using the Eq. (9), we find the 2D electron current density

$$j = j_\omega \exp[\mathrm{i}(hz - \omega t)] + j_+ \exp[i(hz - (\omega + \Omega)t)] + $$
$$+ j_- \exp[i(hz - (\omega - \Omega)t)] + c.c \qquad (10)$$

where the amplitudes are

$$j_\omega = -ieA\frac{(\omega + i\gamma_R)}{D_0}n_0,$$
$$\mathrm{j}_\pm = ieA\alpha\omega_0^2\frac{(\omega + i\gamma_R)n_0}{2D_0}\left(\frac{1}{\Omega_p^2} - \frac{1}{D_\pm}\right) \qquad (11)$$

It is easy to see that, to the first order in the small parameter $\alpha \ll 1$ the reflected and transmitted fields contain two satellites with up- and down-shifted frequencies $\omega \pm \Omega$. in addition to the principal components at the frequency $\omega$. As shown above, the reflected and transmitted principal components propagate at angles to the x-axis, equal to the angle of incidence $\theta$ (see Fig. 1). The up- and down-shifted in frequency satellites, generated by the corresponding current components in Eq. (10), have the same z-component of the wave vector, $h = k \sin\theta$. For the up-shifted satellites, the wavenumber is $k_+ = \sqrt{\varepsilon}(\omega + \Omega) / c$, and thus the scattering angle is

$$\theta_+ = \arcsin\left(\frac{\omega}{\omega + \Omega}\sin\theta\right) \qquad (12)$$

Thus, these satellites propagate closer to the x-axis than the principal components (see Fig. 1). Note that if $\omega \ll \Omega$, the up-shifted components propagate almost normally to the quantum well. The down-shifted in frequency satellites have the wave number $k_- = \sqrt{\varepsilon}|\omega - \Omega| / c$, and it can therefore be concluded that their propagation mode depends heavily on the relationship between the signal and the beat

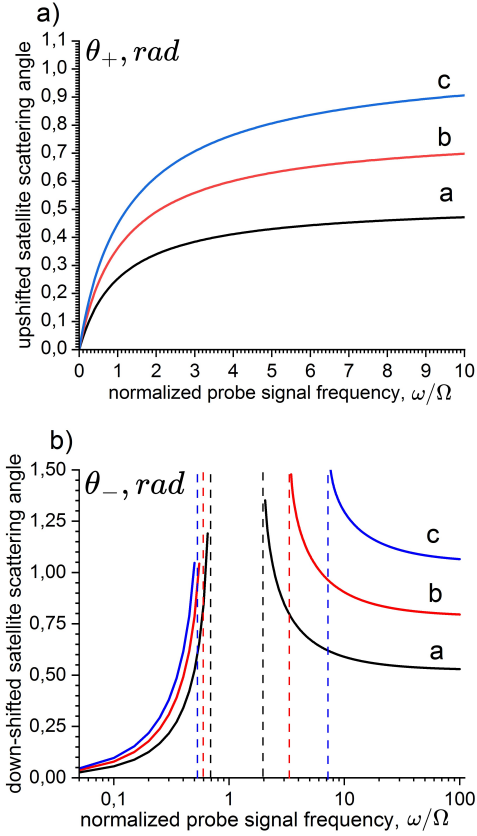

a)

$\theta_+$, rad

c

b

a

upshifted satellite scattering angle

normalized probe signal frequency, $\omega/\Omega$

b)

$\theta_-$, rad

c

b

a

down-shifted satellite scattering angle

normalized probe signal frequency, $\omega/\Omega$

FIG 2. The dependence of reflection and refraction angles of the a) up-shifted ($\theta_+$) and b) down-shifted ($\theta_-$) in frequency satellites vs the probe wave frequency, normalized to the beat frequency, $\omega/\Omega$, at different angles of incidence, (a) $\theta = \pi/6$ (black curves), (b) $\theta = \pi/4$ (red curves), and (c) $\theta = \pi/3$ (blue curves). In b), the dashed vertical lines of the same colour indicate the frequency ranges at which downshifted satellites appear as surface waves; the domain $\omega/\Omega < 1$ corresponds to backward-penetrating downshifted satellites, while the domain $\omega/\Omega > 1$ corresponds to forward propagation.

frequencies. When the signal frequency exceeds the beat frequency so that $\omega(1-\sin\theta) > \Omega$, the down-shifted satellites propagate forward at the angle

$$\vartheta_- = \arcsin\left(\frac{\omega}{|\omega - \Omega|}\sin\theta\right) \qquad (13)$$

i.e., closer to the $z$-axis compared to the principal components. Under the relationship $|\omega - \Omega| < \omega\sin\theta$ the transverse wave number becomes imaginary. This means that both the reflected and transmitted down-

shifted components appear as TM surface waves which propagate forward in the $z$-direction when $\omega > \Omega$, and opposite direction when $\omega < \Omega$. Finally, if $\omega(1+\sin\theta) < \Omega$, then both the reflected and transmitted down-shifted satellites are backward propagating plane waves at the angle given by Eq. (13). The dependences of the angles $\theta_\pm$ on the probe wave frequency normalized to the beat frequency $\omega/\Omega$ are shown in the Fig. 2 at different incidence angles, $\theta = \pi/6, \pi/4, \pi/3$. When $\omega << \Omega$, both satellites propagates almost normally at small angles, and when $\omega >> \Omega$ the up- and down-shifted in frequency satellites propagates close to the principal components. It is interesting to note that at $\omega = \Omega/2$ the backward reflected down-shifted component propagates exactly opposite to the incident plane wave and has the same frequency, i.e., it emerges some kind of the phase conjugation.

The amplitudes of the refracted and transmitted waves are determined by the boundary conditions. That is, the difference in the tangential components of the total magnetic field above ($H_{1t}$) and below ($H_{2t}$) the QW is proportional to the 2D current density:

$$[\hat{e}_x(H_{1t}-H_{2t})] = 4\pi j_{2D}/c \qquad (14)$$

and the continuity in the longitudinal components of the electric field strengths on the surface of 2D QW takes place.

For the principal components, the above boundary conditions give

$$\frac{kc}{\omega}(E_{p0} + E_{p\sigma}^R - E_{p\sigma}^T) = -i\frac{4\pi}{c}\frac{(\omega + i\gamma_R)}{D_0}eAn_0,$$
$$(E_{p0} - E_{p\sigma}^R) = E_{p\sigma}^T \qquad (15)$$

Introducing the dimensionless parameters $\psi = 4\pi n_0 e^2 k/\varepsilon m\omega^2$ and $\eta_p = \omega(\omega + i\gamma_R)/D_0$, we find the following formulas for the amplitudes of the reflected and transmitted principal components

$$\frac{E_{p\sigma}^R}{E_{p0}} = -\frac{i\psi\eta_p\cos\theta}{1 - i\psi\eta_p\cos\theta},$$
$$\frac{E_{p\sigma}^T}{E_{p0}} = \frac{1}{1 - i\psi\eta_p\cos\theta} \qquad (16)$$

*Contact author: smetaniniv@lebedev.ru

Using the same method for satellites that have been shifted up or down in frequency, we find that

$$\frac{E_{p\pm}^R}{E_{p0}} = \frac{E_{p\pm}^T}{E_{p0}} = -\frac{\alpha}{2} \frac{i\psi\eta_p \cos\theta}{1 - i\psi\eta_p \cos\theta} \left( \frac{\omega_0^2}{\Omega_\pm^2} - \frac{\omega_0^2}{D_\pm} \right) \quad (17)$$

In other words, the amplitudes of the satellites differ from that of the unshifted frequency-reflected component by a factor of $(\alpha/2)(\omega_0^2/\Omega_\pm^2 - \omega_0^2/D_\pm)$.

## IV. SCATTERING OF THE S-POLARIZED WAVE

Now, let us consider the case of an s-polarised probe plane wave incident on a laser-driven two-dimensional quantum well (2D QW). The electric field strength of this wave is perpendicular to the plane of incidence

$$\vec{E}_s = \hat{e}_y E_{s0} \exp[i(kz \sin\theta - kx\cos\theta - \omega t)] + c.c. \quad (18)$$

From the hydrodynamic model [29], it is clear that s-polarised fields do not excite 2D electron density oscillations in the QW, nor the longitudinal self-consistent electric field. All reflected and transmitted waves, including principal and up- and down-shifted satellites, are also s-polarised. Consequently, the current density in the QW plane has only the y-component and oscillates as follows

$$j_s = j_{s\omega} \exp[i(hz - \omega t)] +$$
$$+ j_{s+} \exp[i(hz - (\omega + \Omega)t)] + \quad (19)$$
$$+ j_{s-} \exp[i(hz - (\omega - \Omega)t)] + c.c.$$

with the amplitudes to the first order in $\alpha$ given by

$$j_{s\omega} = -i\frac{e^2 n_0}{m} \frac{E_{s0} + E_{s\omega}^R + E_{s\omega}^T}{\omega + i\gamma},$$
$$j_{s\pm} = -i\frac{e^2 n_0}{m} \left\{ \frac{E_{s\pm}^R + E_{s\pm}^T}{\omega \pm \Omega + i\gamma} + \frac{n_{10}}{2n_0} \frac{E_{s0} + E_{s\omega}^R + E_{s\omega}^T}{\omega + i\gamma} \right\} \quad (20)$$

where $E_{s\omega}^R$ and $E_{s\omega}^T$ are the amplitudes of the reflected and transmitted principal components, respectively, and $E_{s\pm}^R, E_{s\pm}^T$ are the amplitudes of the up- and down-shifted in frequency reflected and transmitted waves, respectively. Once again, since all the reflected and transmitted waves have the same z-component of their wave vectors $h = k\sin\theta$, all the above considerations regarding the refraction and transmission angles remain valid for s-polarised waves.

The boundary conditions (14) for the principal components are as follows

$$-\frac{kc}{\omega}\left[ E_{s0} - E_{s\omega}^R - E_{s\omega}^T \right]\cos\theta = \frac{4\pi}{c} j_{s\omega},$$
$$E_{s0} + E_{s\omega}^R = E_{s\omega}^T \quad (21)$$

that lead to the following formulas for their amplitudes ( $\eta_s = \omega/(\omega + i\gamma)$ )

$$\frac{E_{s\omega}^R}{E_{s0}} = -\frac{i\psi\eta_s}{\cos\theta + i\psi\eta_s}, \quad \frac{E_{s\omega}^T}{E_{s0}} = \frac{\cos\theta}{\cos\theta + i\psi\eta_s} \quad (22)$$

For the up- and down-shifted in frequency ( $\omega \pm \Omega$ ) components the boundary conditions are

$$\frac{k_\pm c}{\omega_\pm}\left[ E_{s\pm}^R + E_{s\pm}^T \right]\cos\theta_\pm = \frac{4\pi}{c} j_{s\pm},$$
$$E_{s\pm}^R = E_{s\pm}^T \quad (23)$$

and we find the amplitudes as

$$\frac{E_{s\pm}^R}{E_{s0}} = \frac{E_{s\pm}^T}{E_{s0}} = -\frac{1}{2}\frac{n_{10}}{n_0}\frac{i\psi\eta_s \cos\theta}{\cos\theta + i\psi\eta_s}\frac{1}{\cos\theta_\pm + i\psi\eta_\pm}. \quad (24)$$

Note that in the case of normal incidence, $\theta = 0$, we have $\eta_s = -\eta_p$, the plasma frequency becomes zero, $\Omega_p = 0$, and the field amplitude equations (16) and (22) for the principal components, as well as the equations (17) and (24) for the up- and down-shifted satellites, coincide to first order in perturbation theory for both s- and p-polarised incident waves.

It is interesting to note that in the case of an s-polarized incident wave at frequencies within the domain $\Omega/(1 - \sin\theta) > \omega > \Omega/(1 + \sin\theta)$, the down-shifted satellites represent the TE surface waves, $(H_z, H_x, E_y)\exp(-\kappa_\perp |x|)\exp(i[hz - (\omega - \Omega)t])$, where $\kappa_\perp = (h^2 - \varepsilon(\omega - \Omega)^2/c^2)^{1/2}$. It is well known that the free propagation of TE surface electromagnetic waves is prohibited [31]. However, in our case, we have in the externally driven two-dimensional plasma oscillations in the structure, so that TE surface electromagnetic waves become possible. The amplitude of the electric field strength $E_s^{surf}$ in this surface wave is given by

$$\frac{E_s^{surf}}{E_{s0}} = -\frac{1}{2}\frac{n_{10}}{n_0}\frac{\psi\eta_s\sqrt{\varepsilon}\cos\theta}{\cos\theta + i\psi\eta_s}\frac{(\omega - \Omega)}{\kappa_\perp c + i\psi\eta_-(\omega - \Omega)}. \quad (25)$$

*Contact author: smetaninv@lebedev.ru

## V. CONCLUSIONS

Let us make some estimates assuming a GaAs 2D quantum well [28]. We assume that the relaxation rate are the rate of electron-hole recombination $\gamma_R \sim 10^9 s^{-1}$, and the momentum relaxation rate $\gamma \Box 10^{11} s^{-1}$ at a temperature of 77 K [32]. At driving laser intensities $P_1 = P_2 \approx 10 \text{kW/cm}^2$, the steady-state density of generated electron-hole pairs can be estimated as $n_0 \approx 10^{12} cm^{-2}$ [28]. For GaAs, ($\varepsilon = 12.9$ and the electron mass $m/m_0 = 0.067$), we find $\psi \approx 4.4 \times 10^{11} s^{-1}/\omega$, and the density modulation $n_{10}/n_0 \approx \gamma_R/\Omega$. Assuming both the modulation frequency and the probe signal frequency are of the same order, $\omega \sim \Omega \sim 10^{12} s^{-1}$, we estimate the amplitude of the up- and down-shifted in frequency satellites as $E_{s,p\pm}/E_{s,p0} \sim \psi n_{10}/n_0 \approx 4.4 \times 10^{-4}$ which are quite measurable values. Note that $n_{10}/n_0 \approx \gamma_R/\Omega$ is found above for homogeneous plasma density oscillations, this ratio can be significantly increased when the resonance plasma wave is excited [28], which is more complicated case due to the spatiotemporal structure of 2D plasma oscillations, and we plan to consider it in our future work. Also note that the above consideration is only suitable in a steady-state situation and thus invalid when $\omega \to \Omega$, that is, when the time taken to reach steady state is infinite, or when $|\omega - \Omega| \to \omega \sin \theta$, that is, the characteristic transverse size of the field is infinite. The above consideration makes it possible to conclude that photo-excited by two-color laser radiation 2D QW could be a prospective realization of a photonic time crystal in THz and even optical frequency domains.

## ACKNOWLEDGMENTS

This work has been supported by the Russian Federation State Contract #FFMR-2024-0009.

AUTHOR DECLARATIONS

AUTHOR CONTRIBUTIONS
Both authors (I. S. and A.U.) contributed equally to this work.

CONFLICT OF INTEREST STATEMENT
The authors have no conflicts to disclose.

DATA AVAILABILITY
The data that supports the findings of this study are available within this article.

*Contact author: smetaninv@lebedev.ru